\documentclass[11pt,notitlepage]{article}
\usepackage{amsfonts}
\usepackage[sumlimits]{amsmath}
\usepackage{amsthm}
\usepackage{amssymb}
\usepackage{setspace}
\usepackage{graphicx}

\newcommand{\be}{\begin{eqnarray}}
\newcommand{\ee}{\end{eqnarray}}
\newcommand{\De}{\Delta}

\newcommand{\eps}{\varepsilon}

\newcommand{\ba}[1]{\begin{array}{#1}}
\newcommand{\ea}{\end{array}}

\begin{document}

\begin{flushright}
 \mbox{\normalsize ITEP/TH-42/10}
\end{flushright}

\bigskip

\centerline{\Large{\textbf{A model of conductivity in polymer films with two conductivity states.}}}

\bigskip

\centerline{Andrey D. Vlasov \footnote{Email: vlasov.ad@gmail.com}}

\bigskip

\centerline{\it Institute of theoretical and experimental physics, Moscow, Russia}
\centerline{\it Prokhorov General Physics institute, Moscow, Russia}
\centerline{\it Moscow institute of physics and technology, Dolgoprudnuy, Moscow, Russia}

\bigskip

\begin{abstract}
We suppose and develop a simple quantitative model of polymer film conductivity. This model can be seen as a further development of the ideas of Vlasov, Apresyan et al \cite{CondSwitchPress}. The main point of the model is that conducting islands exist, and the charge transfer between the islands is carried out by mobile segments of polymer molecules. This model quantitatively describes the presence of two states of conductivity, and the current stabilization phenomena, and it predicts the temperature dependence of conductivity, and the dependence of conductivity on the thickness of the polymer film. The pressure-driven transition to high-conductivity is described only qualitatively in this model.
\end{abstract}

\tableofcontents

\section{Introduction}
\subsection{Motivation}
In recent years interest in the conductivity of polymer films has greatly increased \cite{Lachinov},\cite{Yang} , and several Nobel prizes in physics and chemistry were awarded in recognition of people's success in polymer film conductivity. There is a plethora of possible applications of polymer electronics in different fields: in science, in technical devices, in computers, etc. But to use polymer electronics we should understand polymer conductivity better, or at least create a model which answers the most simple questions: "Why and how do the polymers conduct current? What are the two states of conductivity? How can we switch between them and why? Why does the stabilization of current occur?" and so on; there are a lot of questions. This paper pretends to be a toy model of polymer conductivity, which collects and explains all notable facts known about the conductivity in plasticized PVC (polyvynilchloride): the two states of conductivity and quasi-spontaneous transitions between them \cite{switchlit}, the current stabilization in the state of low conductivity \cite{nonlin}, the pressure-caused transition to a high conductivity state, and the relaxation of current upon the constant voltage. In addition, the model gives the quantitative answers to all the above-mentioned questions, excepting the pressure-induced transition to high conductivity. The pressure-induced transition is explained only qualitatively in this model. This model was developed by using the facts about the conductivity of PVC films, but it is supposed to catch some general patterns of a great variety of different polymers. 
\subsection{Outline of the model}
\begin{figure}[!h]
\begin{center}
\includegraphics[width=0.5\linewidth]{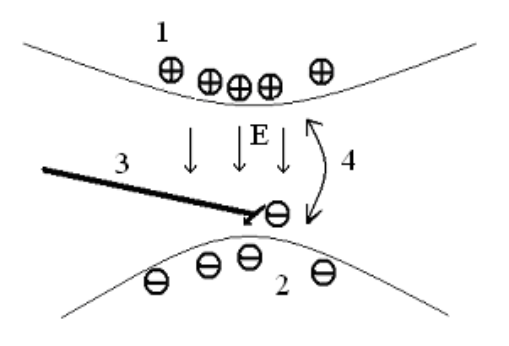}
\caption{The scheme of charge transfer between islands. 1,2 - conducting islands, 3 - moving molecule segment, 4 -    the movement of the segment}
\label{segment}
\end{center}
\end{figure}

Our model is a development of the ideas of D. V. Vlasov, L. A. Apresyan et al \cite{CondSwitchPress}. The main ideas are the following. The structure of polymers is known to be very complicated and involute, sometimes exhibiting fractal behaviour. So, the density in polymers is not a constant, but it varies from place to place and depends on the characteristic size on which we calculate the density. In the regions of high density the atoms of different molecules can approach each other so closely that the wave functions of electrons begin to smear between two or more different atoms. In such regions the delocalisation of electrons can occur and thus the electrons can move within such an area almost freely. This area of a polymer may be similar by its conducting properties to semiconductor. Further we call such areas "conducting islands". The only remaining question is - how is the charge transferred between the different conducting islands? In this model the moving segments of polymers molecules are believed to do it, see Fig.\ref{segment}. 
Without plasticizers the PVC molecules are like rigid sticks, but plasticizers allow them to bend. Some of such segments are within reach of two islands, as shown in the picture Fig.\ref{segment}. The sequence of conducting islands, connected with such junctions, we call the conducting channel. The segment can have a defect, which can hold some charge, most probably several unit charges. This defect can carry both signs of charge, but a preferred sign may exist. If so, there are two modes of conductivity. In the first mode, the segment moves in one direction with a charge, but in the opposite direction without a charge, only due to temperature fluctuations. Furthermore, the second mode is where the segment moves with a charge in both directions. When the segment moves with a charge, it moves upon action of the electric field, and therefore it moves faster, given that field is more than some critical field. These two modes of conductivity are the two conductivity states, and if the movement of a segment upon the field is faster than the movement upon the temperature fluctuations, the current in low-conductivity state is roughly constant with applied voltage, as it was observed in experiments. The transitions occur upon the high voltages, and upon suddenly supplied voltages.
\subsubsection{Pressure-induced phenomena}
 The pressure can have the same consequences as the applied field: due to the pressure, the neighbouring conducting islands become closer, and the field between them increases. In addition, the pressure causes the charge division even without the external field. This is because the reaction force which appears when the pressure is supplied is an electromagnetic force (not gravitational, weak or strong), hence, the charges should exist to produce electric force. This charge division in polymer materials can be easily observed: your plastic comb attract small pieces of paper after you have used it. The appearance of the macroscopic force tells us that huge charges play a role in this phenomena - this experiment can be used to reinforce the argument that pressure causes great charge division. On the other hand, the applied field causes the charge division, and at different sides of each junction are the charges of different signs. These charges attract each other, creating the pressure in the junction. 
\section{The description of the model \label{descr}}
\subsection{The dependence on the thickness of the polymer}
In our model, it is supposed that the current goes by some channels which are nothing but the system of islands, connected with a conducting junction. Indeed, not any junction can conduct current. To conduct current, junction must have an appropriate mobile segment, and this mobile segment should be appropriately oriented. The existence of mobile segment itself require the existence molecule of plasticizer in junction, because it is the plasticizer molecule which enables the PVC molecule to be not rigid. From this we can infer, that the more the concentration of plasticizer is, the more the conductivity is. More detailed consideration is postponed to Sect.\ref{plast}. Additionally, these molecules should be properly oriented. From all these considerations we can infer, that in our model not each junction can conduct electricity, but electricity flows by some sequences of islands and conducting junctions, which we call conducting channels.
But the idea of channels itself can tell us a lot of information about the dependence of conductivity on thickness. Let's denote the average number of conducting channels per unit area in the polymer as $f(D)$, where $D$ is a thickness of polymer film.
Now we can consider two adjacent films, or, better, one film divided by plane, parallel to the sides of the film. The two  layer of the polymer have thickness of $D_1$ and $D_2$ correspondingly, $D_1+D_2=D$.  Denote the size of conducting island $d$, and consider the piece of the polymer of area $S$. We assume here that the width of channel equals to the size of conducting island. There are $Sf(d_1)$ channels from the upper layer, and $Sf(d_2)$ channels of bottom layer, but the number of channels which go through all composite film is lower, because not all channels of upper layer meet channels of bottom layer. Now we estimate this number, assuming that 1) the channels are rare 2) they are independent. Consider some particular channel of upper layer. To meet with it, the center of bottom layer channel must fall into the circle of radius $d$ around the center of the channel under consideration. This region has an area of $\pi d^2$. Thus, the number of bottom-layer channels falling in this region is $\pi d^2f(D_2)$. So, one channel of upper layer increases the number of channels through entire film by $\pi d^2 f(D_2)$, and there are $S f(D_1)$ upper layer channels. So, the number of channels in entire film equals:
\be
 Sf(D)=\pi d^2f(D_1)Sf(D_2) \\
 f(D_1+D_2)=\pi d^2 f(D_1)f(D_2)
\ee
This functional equation on $f(D)$ has a following solution:
\be
 f(D)=\frac{1}{\pi d^2}\exp{\left(-\frac{D}{d_0}\right)}
\ee
The only arbitrary constant here is $d_0$, the coefficient in front of exponent is fixed by the equation. Of course, this consideration is rather robust, it neglects that the channels are of different sizes, that they are not independently distributed over area, the layers may not be independent, the edge effects etc.
\subsection{The one-channel current}
\begin{figure}[h!]
\begin{center}
\includegraphics[width=0.5\linewidth]{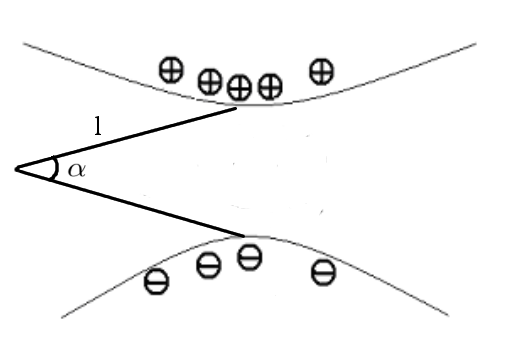}
\caption{The parameters $l$ and $\alpha$ are depicted on this figure.}
\label{lalpha}
\end{center}
\end{figure}

In our model, there are conducting islands and the charge is transferred between different islands by moving segments. Let $d$ be the size of conducting islands and $\sigma_0$ be their conductivity. Of course, both the size, conductivity and other parameters are different for different islands. $d$ and $\sigma_0$ represent the average values of these parameters, and this simple model may catch the general patterns of the conductivity. Denote the length of the mobile segment as $l$, and an angle it turns by as $\alpha$ (see Fig.\ref{lalpha}).
The linear density of mobile segment is $\rho_l$ - but it is not a free parameter of the model: it is determined by the structure of polymer molecule. Denote the dimensionless parameter $\gamma\equiv d/(2\sin(\alpha/2)l)$ -this is a ratio of the size of islands to the gap between them.
The current through one junction equals:
\be
 I_{1junc}=\frac{e}{\De t_T+\De t_E} {\rm \quad -\quad low\quad conductivity} \label{jgenT} \\
 I_{1junc}=\frac{2e}{2\De t_E}=\frac{e}{\De t_E} {\rm \quad -\quad high\quad conductivity} \label{jgenE}
\ee
where $\De t_T$ is time of moving of mobile segment from one island to another upon temperature fluctuations, $\De t_E$ is a time of moving of mobile segment from one island to another upon the field acting on the carried charge.   The current flowing through one island equals:
\be
 I_{1island}=E_{d}\sigma_0 d^2
\ee
where $E_d$ is a field inside the conducting islands, and $E_l$ is a field between the islands. $d^2$ in order of magnitude equals to the cross-section of the island. In the stationary case, that is when the charge density does not depend on time everywhere, the $I_{1junc}=I_{1island}$. Then:
\be
I_{1island}=I_{1junc} \label{eqstat} \\
E_d\sigma_0d^2=I_{1junc} \label{fieldsone} \\
E_{mac}(1+\gamma)=\gamma E_d+E_l \label{fieldstwo}
\ee
Here $E_{mac}$ is a macroscopic average field. Once we have found the one-junction current, we can obtain the overall current density:
\be
 j=I_{1island}f(D)
\ee
Now we are going to discuss the dependence of $\De t_T$ and $\De t_E$ on the parameters of model and applied field. Moving segment consists of thousands and possibly millions of atoms, so we can describe it classically. Then, its rotation has the mean kinetic energy equal to $\frac12kT$. So (here $I$ is the momentum of innertia of the segment, $I=\frac13\rho_l l^3$):
\be
 \frac12I\omega^2=\frac12kT \\
 \De t_T=\alpha\sqrt{\frac{\rho_l l^3}{3kT}} \label{dett}
\ee
This formula does not take into account the fact that the motion of the segment is rather chaotic when moving upon the action of temperature fluctuations. At least, it can undergo the collisions with other molecules. If it has collided $N$ times with other molecules, its mean velocity is decreased by $\frac{1}{\sqrt N}$. There are many ways to estimate $N$. By order of magnitude, $N$ may be equal to:
\be
 N=\left(\frac{\alpha l}{l_{con}}\right)^2, \label{Nsquared}
\ee
where $l_{con}$ is length of one connection between atoms in molecule. This estimate can be derived as follows. During the motion the end of segment will get over the path of $\alpha l$. On this path, it will each $l_{con}$ meet with other molecule - so the total number of collisions can be estimated as $\frac{\alpha l}{l_{con}}$. The motion is two-dimensional, so we can take this into account  by squaring the expression. So, the final formula for $\De t_T$ is:
\be
 \De t_T=\alpha\sqrt{\frac{N\rho_l l^3}{3kT}}
\ee
 In order to obtain the expression for $\De t_E$ we use the equation for angular momentum ($I$ here means the moment of inertia of the segment, $I=\frac13\rho_ll^3$):
\be
 I \dot{\omega}=E_lel\frac{1+\cos(\alpha/2)}{2}=eE_ll \cos\left(\frac{\alpha}{4}\right) \\
 \frac12\dot{\omega}\De t_E^2=\alpha \\
 \De t_E=\sqrt{\frac{2\alpha\rho_l l^2}{3eE_l\cos\left(\frac{\alpha}{4}\right)}} \label{dete}
\ee
Since the initial velocity is supposed to be zero, the time of movement is in inverse proportion to the square root of acceleration, i. e. to the square root electric field between islands, $E_l$. Note however, that this formula is true when the velocity obtained due to temperature fluctuations is negligible comparing with the velocity obtained due to the electric field, that is, $\De t_T \gg \De t_E$. Otherwise, temperature fluctuations play sufficient role even when moving upon the field, and this can affect the resulting time of movement.
\subsubsection{Voltage-current curve in high conductivity  \label{UIE}}
In this case, the current through one junction is determined by the formula (\ref{jgenE}), it equals:
\be
 I_1=\frac{2e}{2\De t_E}=\frac{e}{\De t_E} 
\ee
We write equations (\ref{fieldsone},\ref{fieldstwo}) in this case:
\be
E_{mac}(1+\gamma)=\gamma E_d+E_l \\
\sigma_0 E_d d^2=\frac{e}{\De t_E}
\ee
We introduce here the new derived parameter, $E_0\equiv\frac{3e^3\cos\left(\frac{\alpha}{4}\right)}{2\alpha\sigma_0^2d^4 \rho_l l^2}$. Using the $E_0$, one can obtain such an equation and the solutions of this equation:
\be
e'_{mac}=x+x^2; \quad e'_{mac}=\frac{(1+\gamma)E_{mac}}{\gamma^2E_0};\quad x=\sqrt{\frac{E_l}{\gamma^2E_0}} \\
e'_{mac}=\frac{E_{mac}}{E_{maccritE}};\quad E_{maccritE}=\frac{\gamma^2}{1+\gamma}E_0 \\
x=\frac{-1\pm\sqrt{1+4e'_{mac}}}{2}=\frac12(\sqrt{1+4e'_{mac}}-1)\approx \left\{\ba{ll}e'_{mac} &,e'_{mac} \ll 1 \\
\sqrt{e'_{mac}}&,e'_{mac}\gg 1 \ea\right.
\ee
Let's consider first the case when $E_{mac} \ll E_{maccritE}$:
\be
x=e'_{mac} \\
E_{l}=\frac{(1+\gamma)^2}{\gamma^2}\frac{E_{mac}^2}{E_0} \\
I_1=\sigma_0 d^2 \frac{1+\gamma}{\gamma}E_{mac} \\
j=\sigma_0 \frac{\gamma+1}{\gamma\pi}\exp{\left(-\frac{D}{d_0}\right)}E_{mac} \label{jEone}
\ee
In this limit case, the current is proportional to the applied field. The overall conductivity in this case
\be 
\sigma'=\sigma_0\frac{\gamma+1}{\gamma\pi}\exp{\left(-\frac{D}{d_0}\right)}
\ee
In other limit case, $E_{mac}\gg E_{maccrtE}$:
\be
 x=\sqrt{e'_{mac}} \\
 E_l=(\gamma+1)E_{mac} \\
 I_1=\sigma_0 d^2\sqrt{(1+\gamma)E_0E_{mac}} \\
 j=\frac{\sigma_0}{\pi}\sqrt{(1+\gamma)E_0E_{mac}}\exp{\left(-\frac{D}{d_0}\right)} \label{jEtwo}
\ee
In this case, the current is proportional to the square root of the applied voltage. This has very simple physical interpretation: the current in conducting islands is proportional to applied field, and the current through junctions is proportional to the square root of applied field. These "resistors" are connected consequently, so the overall current is determined by the lowest current. Upon low values of field, the linear function is smaller, and upon the large values of field, the square root of field is smaller. 
\subsubsection{Voltage-current curve in low conductivity \label{UIT}}
Low conductivity state is a state where the mobile segment moves in one direction due to the temperature fluctuations, and in other direction it moves upon the force exerted by $E_l$. So, we substitute in the expression (\ref{jgenT}) the formulas (\ref{dett},\ref{dete}):
\be
E_d=\frac{1}{\sigma_0 d^2}\frac{e}{\De t_T+\sqrt{\frac{2\alpha \rho_l l^2}{3eE_l\cos\left(\frac\alpha4\right)}}} \label{eqEmEdElT}
\ee
We denote 
\be 
\De t_0=\frac{2\alpha\sigma_0d^2\rho_l l^2}{3e^2\cos\left(\frac{\alpha}{4}\right)} \\
\eps\equiv\frac{\gamma\De t_T}{\De t_0} \\
x=\sqrt{\frac{E_l}{\gamma^2E_0}}
\ee
and obtain the equation:
\be
 \eps x^3+x^2+x(1-\eps e'_{mac})-e'_{mac}=0; \quad e'_{mac}=(1+\gamma)\frac{E_{mac}}{\gamma^2E_0}
\ee
Let's solve this equation in two limits. The first limit we consider is $\eps e'_{mac} \gg 1$. Then the equation becomes quadratic:
\be
\eps e'_{mac}=1+\eps x^2 \\
x=\sqrt{e'_{mac}-\frac{1}{\eps}}=\sqrt{e'_{mac}}\\
E_l=E_{mac}(1+\gamma) \\
\frac{\De t_T}{\De t_E}=\sqrt{1+\gamma}\gamma\eps e'_{mac}\sqrt{\frac{E_0}{E_{mac}}}
\ee
Usually, $\gamma>1$. So, when $\frac{E_0}{E_{mac}}>1$, $\De t_T \gg \De t_E$ and the current does not depend on the applied voltage - and this indeed is a stabilization of current, which were discovered in the work \cite{nonlin}. In this limiting case,
\be
I_{1junc}=\frac{e}{\De t_T}=\frac{e}{\alpha}\sqrt{\frac{3kT}{N\rho_l l^3}} \\
j_T=I_{1junc}f(D)=\frac{e}{\alpha\pi d^2}\sqrt{\frac{3kT}{N\rho_l l^3}}\exp{\left(-\frac{D}{d_0}\right)} \label{jT}
\ee
The nearly constant current in this case is proportional to $\sqrt{T}$ - this can be verified by experiments. From the formulas (\ref{jgenE},\ref{jgenT}) one can derive, that $\frac{\De t_T}{\De t_E}=\frac{j_{high}}{j_{low}}$, where the $j_{high}$ is a current in high-conductivity state, and $j_{low}$ is the current in low-conductivity state - so in this limit, the currents in the high and low conductivity states have different orders of magnitude. The opposite limit, $\eps e'_{mac}\ll 1$, corresponds to the case when $\De t_T \ll \De t_E$ - and, as we have already mentioned, this case requires more accurate treatment. In this case the time of moving upon the field is not equal to $\De t_E$, defined by (\ref{dete}) - in this process we should take into account the temperature fluctuations when moving upon the field. Let's calculate the limiting field $E_{mac}$, above which $\De t_E \ll \De t_T$. To estimate this, we can formally apply the formula (\ref{jgenT}) in the opposite limit case, $\eps e'_{mac}\ll 1$. In this limit, $\De t_E\gg \De t_T$ and the formula (\ref{jgenT}) reduces to the formula (\ref{jgenE}). So, we obtain that
\be
I_1=\sigma_0 d^2 \frac{1+\gamma}{\gamma}E_{mac}
\ee
We obtain the critical field if we will write $I_1=I_{1T}$. The critical field value equals:
\be
 E_{maccritT}=\frac{\gamma}{1+\gamma}\frac{e}{\alpha\sigma_0 d^2}\sqrt{\frac{3kT}{N\rho_l l^3}}
\ee

\subsection{Transitions between the conductivity states}

\begin{figure}[h!]
\begin{center}
\includegraphics[width=0.6\linewidth]{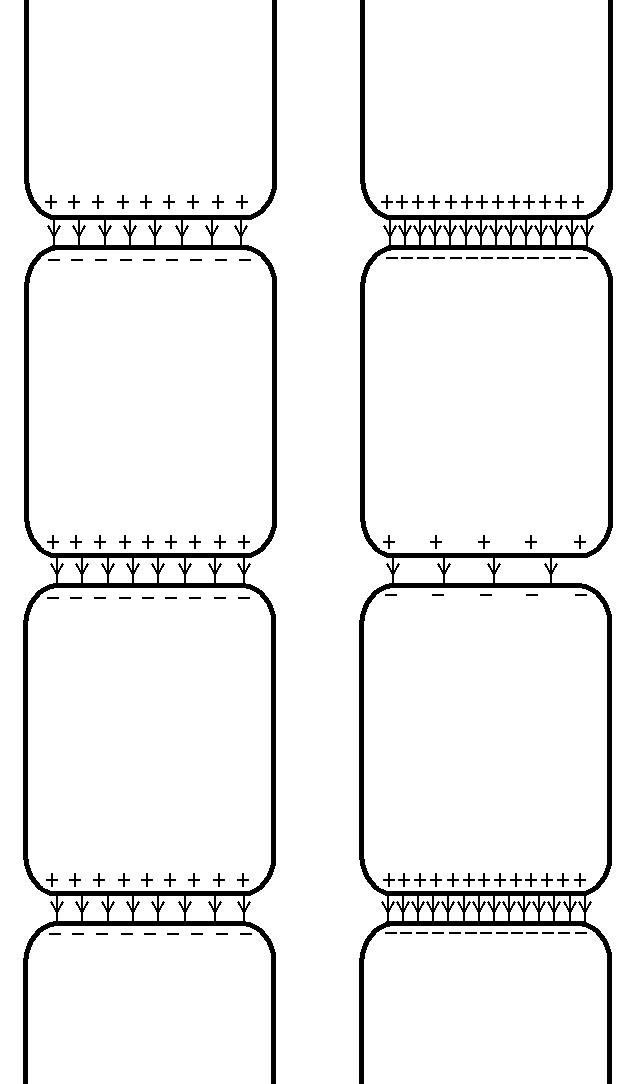}
\caption{The scheme of conducting islands. The left picture is before the transition of middle junction in high-conductivity state, and the right is after such a transition. After transition the field in the same junction falls in magnitude, but in neighbouring junctions is increases.}
\label{islandspict}
\end{center}
\end{figure}

The low conductivity state is the state where the segment moves in one direction upon the field, and in other direction upon the temperature fluctuations. The high conductivity state is the state where the segment moves in both directions with carried charge, in both directions upon the field. How and why the transitions between these two states occur? We can name the two reasons for that. The first reason was already mentioned in Introduction. When the external field is applied, the charges inside the conducting islands are divided, and they are found at the edges of conducting islands, creating the field in junction. The charges of different sign are found at the opposite sides of junction - and they attract each other, creating pressure in the junction. Due to this pressure, the charge trapping of both signs becomes easier. This also explains, why the effect of applied voltage is partially equivalent to the effect of external pressure - pressure can also drive the polymer to high-conductivity state. The second reason is that upon big field, the segment with charge acquires big velocity moving through the junction. With this big velocity, it will collide strongly with the conducting island. So the more the field is, the stronger the interaction between the segment and the island is and the higher the possibility to catch the charge is. Catching charge in both mechanisms is a random process which has sufficiently non-zero probability in wide range of applied fields, and in both mechanisms the probability of catching increases with the field. We can consider some critical field - if the field between islands is greater than the critical, the segment will inevitably catch the charge. So, when the field in junctions is less than critical, both states can exist, but if the field is above critical field, only high-conductivity state can exist.
At the same time, the factor exist which makes both states of low conductivity and of high conductivity more stable. Consider the polymer which is in low-conductivity state. This however does not mean that the field in all junctions is below critical field. Assume the field between some two conducting islands is larger than the critical field, then this junction makes a transition in high-conductance state. This transition changes the conductivity of junction by orders of magnitude, and this causes the rearrangement of field and charge patterns, this is shown at the Fig.\ref{islandspict}. One can see, that after transition to high-conductivity state in one junction, the field in this junction is decreased, and in neighbouring junctions is increased. This may have different consequences. The neighbouring junctions may also make a transition to high-conductivity state, and they will increase the field in the next junctions, and this causes them to go in high-conductivity state - eventually the entire polymer will go in high-conductivity state. But this excitation may be dumped and may die without macroscopic consequences. This may occur because after the junction go to high-conductivity state, the field in this junction is decreased, and this junction will tend to go back, in low-conductivity state.
Due to the same reasons, if all junctions are in high-conductivity state, the transition of few of them in low conductivity is unstable, because their conductivity falls, and the field in these junctions increases, causing them to go back in high-conductivity. So, we have shown, that both low-conductivity and high-conductivity states of entire polymer are locally stable, i. e. are stable to transition of few junctions in other state.
So, in some range of supplied voltage the both states of conductivity can exist in polymer, and the transitions between them are rather stochastic. The lower bound of this range is the voltage, upon which in low-conductivity state the field between {\bf some} islands is more than critical field. The higher bound of this range is the voltage, upon which in low-conductivity state the field between {\bf all}\footnote{Or, at least, between the great majority} islands is greater that critical. If the voltage is greater that the just-named upper bound, then the low-conductivity state is globally unstable: if it would exist some time, it would instantly go to high-conductivity state, because the field between all islands is greater that the critical field. The region of voltages where both states can exist is extended by VI characteristic of voltage supplier. In the great majority of cases, the supplied voltage is decreased as the current is increased. This enlarges the instability: if the high-conductivity state suddenly appears, the current is increased, the voltage falls down, and this encourages the polymer to go back in low-conductivity. \\
\subsection{Non-stationary case}
All the above analysis has been made in the stationary case. The consideration of non-stationary phenomena can extend the range of voltages where both states can exist even more. Consider, for example, the suddenly applied voltage. Since the conductivity of conducting islands is higher, at the first moment the charges are divided inside them. In any non-stationary case (including this) the equation (\ref{eqstat}) breaks down, in this particular case at the first moment $I_{1island}>I_{1junc}$. This creates big charges at the sides of islands, and the $E_l$ value can be much more that $E_l$ value calculated using stationary formulas. This $E_l$ can become greater that $E_{lcrit}$, so the transition to high-conductivity state can occur with very low voltages, but supplied suddenly. The sudden charge dividing in the islands just adjacent to the electrodes creates an addition to the current, which looks like a falling exponent and thus it disappears after some time - and very similar relaxation processes were observed. The characteristic time of this process is nothing else that the $RC$ time of a single island, whereas the value of the current itself depends on the concentration of islands. $RC$-time of conducting island equals $\tau=RC=\frac{\eps_0}{\sigma_0}$, where $\eps_0$ is the electricity constant. The density of current in single island equals $j_1=\sigma_0E_{mac}\exp{\left(-\frac{t}{\tau}\right)}$, so the macroscopic mean density of current is
\be 
j_{mac}=nd^3\sigma_0E_{mac}\exp{\left(-\frac{t}{\tau}\right)}
\ee
where $E_{mac}$ is the macroscopic field and $n$ is a concentration (unit per cubic meter) of conducting islands.
All the islands adjacent to the metal electrode contribute to the falling addition of current, not only those who are beginning of conducting channels. In the islands, which are not the beginnings of conducting channels, the established electric field equals to $E_{mac}$, so it appears in this formula.
\subsection{The dependence on the plasticizer concentration \label{plast}}
The concrete form of this dependence can be modelled by some computer experiments. These computer experiments can give us hints about the structure of the polymer. Assume that the concentration and arrangement of conducting islands do not depend on the concentration of plasticizer. Then the number of conducting channels is determined only by number of "percolations", that is, the number of paths going through conducting islands, connected with each other by appropriate junctions (that is, the junction with the plasticizer molecule and with appropriate oriented PVC molecule). The concentration of plasticizer determines the probability of two neighbouring conducting islands to be connected by appropriate junction. The field along the percolation is approximately the voltage divided by the length of percolation. Some answers about the dependence of the number of percolations on the probability of existence of appropriate junction between two neighbouring islands are given by the percolation theory, but in this case what do we need is the number of percolations. In different conducting states, we may need a number of percolations weighted with different functions of percolation length. For example, in low-conductivity state, the one-channel current does not depend on the applied field if it is above $E_{maccritT}$, so we should take into account all paths with the length less than some fixed length. In high-conductivity, we should sum over all percolations with weight of inverse length of percolation, etc. So, strictly speaking, the dependencies of conductivity on thickness in low-conductivity and in high-conductivity states are different. These dependencies can be calculated in computer models which address percolations.   
\section{Conclusion}
We propose a quantitative model of polymer conductivity which describes a majority of observed properties of conductivity in the PVC films: two states of conductivity, spontaneous transitions between them, the stabilization of current in low conductivity state, the relaxation of current upon the constant voltage, the pressure-induced transition to the high-conductivity state. The numerical expressions for most of these phenomena are obtained, except for the pressure. Pressure-induced phenomena are described only qualitatively in this model. The two mechanisms of charge transfer play a role in our model: the charge transfer inside the areas where the delocalisation of electrons occurs, the so-called conducting islands, and the charge transfer by mobile segments of PVC molecules, which were made non-rigid by plasticizer molecules. The conductivity inside the conducting islands is described phenomenologically by introducing their conductivity $\sigma_0$. The charge transfer between the islands is described as the charge transfer by the mobile segment. In the low conductivity state, the mobile segment catches the charge of one sign and moves upon the field, and it moves in another direction upon the temperature fluctuations. In the high conductivity state, the segment moves upon the field in both directions, because it carries the charges of both signs. All quantitative formulas are found in the Sect.\ref{descr}, but here we collect all the most important dependencies of experiment-observed quantities. The stabilized current has the following dependencies:
\begin{enumerate}
\item $j_T\propto\sqrt{T}$
\item $j_T\propto \exp{\left(-\frac{D}{d_0}\right)}$
\item $j_T$ increases with the plasticizer concentration
\end{enumerate} 
There are some directions for the future research. At first, all these formulas are to be compared with experimental data. Our model has five parameters: $l,d,\sigma_0,d_0,\alpha$, but predicts a lot of quantities. Secondly, the non-stationary phenomena require more accurate quantitative treatment. For example, the theory should predict the period and spectral composition of relaxational oscillations \cite{condchange}, which are generated by the polymer film. The dependence of frequency and spectrum of these oscillations can provide a lot of data to be fitted with these five parameters. Another direction of possible research is the extending of considerations of Sect.\ref{plast}: the number of percolations, the weighted number of percolation, the dependence of percolation number on the probability of single junction to conduct - all these quantities can be obtained from the numerical experiments.  
\subsection{The brief summary of formulas}

\begin{figure}[h!]
\begin{center}
\includegraphics[width=0.6\linewidth]{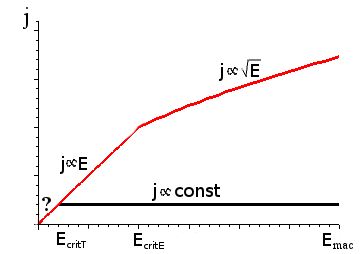}
\caption{A sketch of dependence of $j$ on $E_{mac}$, which is proportional to the supplied voltage. The red curve is for high-conductivity, the black line is for low conductivity. The region where $E_{mac}<E_{maccritT}$ and hence $\De t_E > \De t_T$ is marked by question sign, because this case requires separate treatment.}
\label{graphicspict}
\end{center}
\end{figure}
The formula for the stabilized current in low conductivity, $j_T$, is (\ref{jT}):
\be
j_T=\frac{e}{\alpha\pi d^2}\sqrt{\frac{3kT}{N\rho_l l^3}}\exp{\left(-\frac{D}{d_0}\right)}
\ee
But if the $E_{mac} \ll E_{maccritT}$, this model does not give answer in both states of conductivity, because in this case we should somehow make an estimate of the time of movement of mobile segment upon both temperature fluctuations and small electric force acting on it. The formulas for current in high conductivity are (\ref{jEone}),(\ref{jEtwo}) depending on the relation between $E_{mac}$ and $E_{maccritE}\equiv \frac{\gamma^2}{1+\gamma}E_0$:
\be  
 j_E=\left\{ \ba{rr} \sigma_0 \frac{\gamma+1}{\gamma\pi}\exp{\left(-\frac{D}{d_0}\right)}E_{mac}, & E_{mac} \ll E_{maccritE} \\
 \frac{\sigma_0}{\pi}\sqrt{(1+\gamma)E_0E_{mac}}\exp{\left(-\frac{D}{d_0}\right)}, & E_{mac} \gg E_{maccritE} \ea\right.
\ee
All these relations are schematically shown at the Fig.\ref{graphicspict}.

\section{Acknowledgements}
Author thanks D. V. Vlasov and L. A. Apresyan for fruitful discussions. This work was supported by RFBR grant 09-02-00393, by program of Russian science and education ministry program "Scientific and pedagogical people of Russia" contract 14.740.11.0347 and by Dynasty foundation.

\end{document}